\documentclass[12pt]{iopart}

\usepackage{iopams}
\expandafter\let\csname equation*\endcsname\relax
\expandafter\let\csname endequation*\endcsname\relax
\usepackage{amsmath,mathtools}
\usepackage[colorlinks, linkcolor=black, urlcolor=blue, citecolor=blue, anchorcolor=blue]{hyperref}
\usepackage{soul}
\usepackage{slashed}
\usepackage{cite}
\usepackage{bm}
\usepackage{amssymb}
\usepackage{graphicx}
\usepackage{array}
\usepackage{tabulary}
\usepackage{etoolbox}
\apptocmd{\sloppy}{\hbadness 10000\relax}{}{}
\usepackage{appendix}
\usepackage{amsfonts}
\usepackage{tikz}
\usetikzlibrary{shapes,arrows,positioning}
\usetikzlibrary{decorations.pathmorphing} 
\usetikzlibrary{arrows.meta} 
\usepackage[utf8]{inputenc}
\usepackage{cancel}
\usepackage{varwidth}
\usepackage{xcolor}
\begin{document}
\title[$X(6200)$ as a molecular di-meson state]{An interpretation of the fully-charmed scalar state $X(6200)$ as a molecular di-meson}
\author{Ö.~E.~Demircan}	
\address{Department of Physics, Faculty of Engineering and Natural Sciences, Bursa Technical University, 16310, Bursa, T\"urkiye}
\author{H.~Da\u{g}\footnote{Corresponding author: H.~Da\u{g}, huseyin.dag@btu.edu.tr}}
\address{Department of Physics, Faculty of Engineering and Natural Sciences, Bursa Technical University, 16310, Bursa, T\"urkiye}
\author{H.~Sundu}
\address{Department of Physics Engineering, Faculty of Engineering and Natural Sciences, Istanbul Medeniyet University, 34700, Istanbul, T\"urkiye}
\author{J.~Y.~S\"{u}ng\"{u}}
\address{Department of Physics, Faculty of Arts and Sciences, Kocaeli University, 41001, Kocaeli, T\"urkiye}	
\author{E.~V.~Veliev}
\address{Department of Physics, Faculty of Arts and Sciences, Kocaeli University, 41001, Kocaeli, T\"urkiye}

\vspace{10pt}
\begin{abstract}
The LHCb Collaboration recently observed new structures in the invariant mass spectrum of $J/\psi J/\psi$ meson pairs produced in proton-proton collisions, including a narrow peak around $6.2$ GeV. This study investigates the thermal behavior of this newly discovered fully-charm resonance, assuming it to be an $\eta_c\eta_c$ molecule with quantum numbers $J^{PC} = 0^{++}$. Employing thermal QCD sum rules up to dimension four, we analyze the temperature dependence of the mass and the decay constant. Our results indicate that both physical quantities decrease as the temperature rises. Notably, at zero temperature, the mass and decay constant of the state are consistent with those reported in the existing literature. These findings are expected to provide valuable insights for future experimental investigations in this field.
\end{abstract}
\maketitle
\section{Introduction}
The field of hadron spectroscopy has seen remarkable progress in recent years, particularly with the discovery of numerous exotic hadrons that challenge our understanding of conventional quarks explained by the quark model\cite{Godfrey:2008nc,Guo:2017jvc,Chen:2022asf,Brambilla:2019esw,Albuquerque:2018jkn,Ali:2017jda}. These exotic states, especially those containing heavy quarks, have provided new opportunities to enhance our understanding of the fundamental interactions governed by Quantum Chromodynamics (QCD) in the non-perturbative regime.

The X(6200) resonance is one of the significant recent observations in the exotic hadron spectrum. In 2020, the LHCb Collaboration reported evidence for this fully-charmed state in the di-$J/\psi$ mass spectrum at the center of mass energies of $\sqrt{s} = 7, 8,$ and $13$ TeV, corresponding to an integrated luminosity of $9$ fb$^{-1}$~\cite{LHCb:2020bwg}. This exotic candidate, which may exist either as a molecular state or as a fully-heavy tetraquark, has attracted significant theoretical interest due to its unique position in the spectrum of exotic states. The significance of $X(6200)$ extends beyond its exotic composition, as it provides insights on the binding mechanisms of multiquark states. Its discovery contributes to our understanding of multi-quark systems and strong interaction dynamics at the boundary between perturbative and non-perturbative regimes, potentially revealing new aspects of color confinement and hadron mass generation.

The historical prediction of states with more than three quarks dates back to 1964, when Gell-mann introduced the quark model, where baryons are formed by ann odd number of quarks, and mesons by an even number of quarks \cite{Gell-Mann:1964ewy}. In 1975, Iwasaki predicted the existence of a resonance around $ 6.2 $ GeV, suggesting that it could be an exotic meson with hidden charm \cite{Iwasaki:1975pv}. Since then, various theoretical frameworks have been used to investigate the properties of $X(6200)$, each offering valuable insights into its nature. Among these approaches, the QCD string model describes $X(6200)$ as a compact tetraquark state with quantum numbers $J^{PC} = 0^{++}$, treating it as a bound system of four charm quarks with a predicted mass of 6.196 GeV and a dominant decay into charmonium states~\cite{Nefediev:2021pww}. Similarly, Faustov and his colleagues, using a relativistic quark model, suggest that $X(6200)$ could belong to the fully-heavy tetraquark spectrum, with a possible quantum number assignment of $J^{PC} = 0^{++}$ and a predicted mass of approximately 6190 MeV. Their results also indicate that this resonance may decay into the $\eta_c(1S)\eta_c(1S)$ and $J/\psi(1S)J/\psi(1S)$ channels~\cite{Faustov:2022mvs}.

Furthermore, Dong et al.~\cite{Dong:2020nwy} analyzed the LHCb double-\( J/\psi \) spectrum using a coupled channel approach and found strong evidence for a near-threshold state, identified as \( X(6200) \), with quantum numbers \( J^{PC} = 0^{++} \) or \( 2^{++} \). They suggested that \( X(6200) \) is likely a shallow bound or virtual state in the \( J/\psi J/\psi \) system, possibly indicating a molecular nature. More recently,  a coupled-channel analysis by Song et al.~\cite{Song:2024ykq} hinted at the existence of \( X(6200) \) with a more precise determination of its pole position, finding it in the range \( (6171 - 6202) \) MeV with an imaginary part up to \( (-12) \) MeV, where they suggested that \( X(6200) \) is likely a near-threshold molecular state rather than a compact tetraquark. Additionally, Agaev et al. investigated $X(6200)$ as a fully-charmed hadronic molecule composed of $\eta_c\eta_c$ using QCD sum rules, and determined its mass as $(6264 \pm 50)$ MeV and its total width as $(320 \pm 72)$ MeV, favoring the interpretation of $X(6200)$ as a hadronic molecular state~\cite{Agaev:2023ruu}.

Moreover, the study of fully-heavy bound states at extreme temperatures provides valuable insights into QCD matter under extreme conditions, particularly in high-temperature media like those produced in heavy-ion collisions~\cite{Shuryak:1980tp,Harris:2023tti,Krintiras:2024qzx,Stoecker:2004qu,Elfner:2022iae,Prozorova:2024oal,Torres:2024ile}. Due to their strong binding energies and relatively simple internal structure, fully-heavy molecules serve as sensitive probes for investigating QCD matter at finite temperatures and densities. Their behavior in the quark-gluon plasma (QGP) provides critical information about the medium's temperature, density, and transport coefficients. Such studies may reveal novel phenomena, including the formation of new hadronic states or modifications to existing hadronic properties in hot, dense media. Beyond hadron spectroscopy, this line of research also has significant astrophysical relevance, particularly for the study of neutron stars and supernovae, where matter exists under extremely high densities and temperatures~\cite{Fukushima:2025ujk}. Also, comparisons between experimental findings and theoretical models such as lattice QCD and effective field theories play a crucial role in deepening our understanding of these complex systems. In recent years, numerous theoretical investigations have explored the behavior of hadrons in hot and dense media~\cite{Zhao:2023ucp,Mishra:2023uhx,Sungu:2020zvk,Sungu:2019ybf,Turkan:2019anj,Gungor:2023ksu,Sungu:2020azn,Azizi:2020itk,Azizi:2020yhs,Sungu:2024oax,Veliev:2014tca,Veliev:2011kq,Bozkir:2022lyk,Er:2022cxx}.

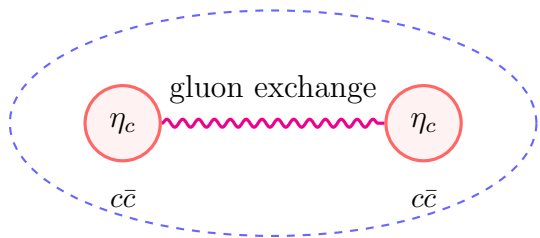
\begin{figure}[htbp]
	\centering 
\begin{tikzpicture}[
	meson/.style={circle, draw=red!60, fill=red!5, very thick, minimum size=10mm},
	gluon/.style={decorate, decoration={snake, amplitude=1.5pt, segment length=6pt}, draw=magenta, very thick},
	binding/.style={dashed, draw=blue!60, thick},
	label/.style={black, font=\small}
	]
	
	\node[meson] (eta1) at (-2, 0) {$\eta_c$};
	\node[meson] (eta2) at (2, 0) {$\eta_c$};
	
	\draw[gluon] (eta1) -- (eta2) node[midway, above=3pt] {$\text{gluon exchange}$};
	
	\draw[binding] (0,0) ellipse (3.5cm and 1.5cm);
	
	\node[label] at (-2, -1) {$c\bar{c}$};
	\node[label] at (2, -1) {$c\bar{c}$};
	
\end{tikzpicture}
\caption{X(6200) ($J^{PC=}0^{++}$) $\eta_c \eta_c$ molecule structure.} 
\label{fig:x6200} 
\end{figure}

In the present work, we explore the thermal properties of the $X(6200)$ state, assuming quantum numbers $J^{PC} = 0^{++}$ and a molecular structure of $\eta_c\eta_c$ as illustrated in Figure~\ref{fig:x6200}. Using Thermal QCD Sum Rules (TQCDSR) up to dimension 4, we analyze the behavior of this exotic hadron under varying temperature conditions, providing new insights into its response to thermal effects.

This paper is structured as follows. Section~\ref{sec:analysis} presents the methodology used to analyze the hadronic properties of $X(6200)$ at finite temperatures, employing TQCDSR up to dimension four. The derivation of the sum rules, the interpolating currents used, and the relevant operator product expansion (OPE) terms are detailed in this section. Then, we discuss the numerical results obtained for the mass and decay constant of $X(6200)$ as a function of temperature, highlighting its behavior near the critical temperature. Finally, Section~\ref{sec:conclusion} summarizes our conclusions and discusses the implications of our results for future experimental investigations of fully-heavy hadronic states in extreme environments.
\section{Analysis of Hadronic Properties of $X(6200)$ Resonance at Finite Temperatures}\label{sec:analysis}
TQCDSR is an extension of traditional QCD sum rules~\cite{SVZ} that incorporates temperature effects, enabling us to study hadronic properties under finite temperature conditions~\cite{Bochkarev:1985ex}. This approach is especially valuable for understanding hadronic matter in extreme environments, such as high-energy collisions or early universe conditions. A key parameter in TQCDSR analysis is the critical temperature ($T_c$), which is the transition temperature from QGP to the hadron phase. While this value can vary based on collision conditions and measurement methodologies, current estimates place it between $150$ and $170$ MeV. In this study, we adopt $T_c = 155$ MeV consistent with recent literature~\cite{Andronic:2017pug,Aoki:2006br,Steinbrecher:2018phh,Fischer:2018sdj}.

The sum rules for temperature dependent spectral parameters of the $X(6200)$ state are constructed from the correlation function

\begin{equation}\label{eq::corr1}
\Pi(p,T) = i \int d^4x e^{ip \cdot x} \langle \Omega | \mathcal{T} \{ J(x) J^\dagger(0) \} | \Omega \rangle,
\end{equation}
where $\Omega$ represents the hot medium and $\mathcal{T}$ is the time-ordering operator. The thermal average is defined as
\begin{equation}
\langle \hat{O} \rangle = \frac{\text{Tr} (e^{-\beta H} \hat{O})}{\text{Tr}(e^{-\beta H})},
\end{equation}
where $\langle \hat{O} \rangle$ is the thermal average of the operator $\hat{O}$. The numerator is the trace (Tr) of the product of the thermal equilibrium state \( e^{-\beta H} \) and the operator \( \hat{O} \), while the denominator is the trace of the thermal equilibrium state alone, representing the sum over all possible quantum states of the system. Here, \( \beta = 1/(k_B T) \), where \( k_B \) is the Boltzmann constant and \( T \) is the temperature.

In order to evaluate the correlation function given in Eqn. \ref{eq::corr1}, the interpolating current is defined as
\begin{eqnarray}
J(x) &=& \bar{c}_a (x)i\gamma_5 c_a (x) \bar{c}_b (x) i\gamma_5  c_b (x).
\end{eqnarray}
where $a$ and $b$ are color indices. This choice reflects the interpretation of the X(6200) resonance as a weakly bound molecular state composed of two pseudoscalar $\eta_c$ mesons, with quantum numbers $J^{PC} = 0^{++}$. Combination of these pseudoscalar currents yields the correct quantum numbers for a scalar state, and the color-singlet structure enables a consistent description of meson-meson interactions near threshold.

The physical (phenomenological) side of the correlation function at finite temperature takes the form
\begin{equation}
\Pi^{\text{Phys.}} (p,T) = \frac{m^2 (T) f^2 (T)}{m^2 (T) - p^2} + \text{(subtracted terms)},
\end{equation}
where \( m(T) \) and \( f(T) \) denote the temperature-dependent mass and coupling of the scalar tetraquark state, respectively. The subtracted terms represent contributions from higher resonances and continuum states, which are first suppressed via Borel transformation and then modeled and removed using the parton-hadron duality assumption within the conventional QCD sum rule framework. The matrix element defining the coupling of the interpolating current to the scalar meson with quantum numbers \( J^{PC} = 0^{++} \) is defined as
\begin{equation}
\langle \Omega | J |X(p) \rangle_T = f (T) m (T).
\end{equation}
In the physical side of the calculations, $\Pi^{\text{Phys.}} (p,T)$ is written as a dispersion integral of the form
\begin{equation}
 \Pi^{Phys}(p^2,T) = \int_{4m^2}^{\infty} \frac{\rho^{Phys}(s,T)}{s - p^2} ds + \dots ,  
\end{equation}
where $\rho^{Phys}(p^2,T)$ is the imaginary part of $\Pi^{Phys}(p^2,T)$, and the infinite integral includes all states coupling to selected interpolating current.

On the other hand, the correlation function in Eqn. \ref{eq::corr1} is also calculated by inserting the interpolating currents and then contracting the quark fields using Wick’s theorem. This results in an expression involving traces over the products of heavy quark propagators in coordinate space at finite temperature, which can be written as 
\begin{align}
	\nonumber	\Pi^{\text{QCD}}(p, T) = i \int d^4x\, e^{ip \cdot x} \Big\{ 
	\nonumber	& \text{Tr}[\gamma_5 S_c^{b a'}(x) \gamma_5 S_c^{a' b}(-x)] \, \text{Tr}[\gamma_5 S_c^{a b'}(x) \gamma_5 S_c^{b' a}(-x)] \\
	\nonumber	& - \text{Tr}[\gamma_5 S_c^{b b'}(x) \gamma_5 S_c^{b' a}(-x)] \, [\gamma_5 S_c^{a a'}(x) \gamma_5 S_c^{a' b}(-x)] \\
	\nonumber	& - \text{Tr}[\gamma_5 S_c^{b a'}(-x) \gamma_5 S_c^{a' a}(-x) \gamma_5 S_c^{a b'}(x) \gamma_5 S_c^{b' b}(-x)] \\
		& + \text{Tr}[\gamma_5 S_c^{b b'}(x) \gamma_5 S_c^{b'b}(-x)] \, \text{Tr}[\gamma_5 S_c^{a a'}(x) \gamma_5 S_c^{a'a}(-x)] 
		\Big\}_T.
	\label{eq::PQCD1}\end{align}
Here: 
\begin{itemize}
	\item $a, b, a', b'$ are color indices running from 1 to 3 in QCD,
	\item $\gamma_5$ is the Dirac gamma matrix appearing in pseudoscalar currents,
	\item $\text{Tr}$ stands for the trace over Dirac gamma matrices,
	\item The subscript $T$ indicates that the correlation function is evaluated at finite temperature,
    \item $S_c^{ij}(x)$ denotes the heavy (charm) quark propagator, with $i$ and $j$ being color indices, and given by
\end{itemize}
\begin{align}
S_c^{ij}(x) &= i \int \frac{d^4 p}{(2\pi)^4} e^{-ip \cdot x} \Bigg[ \frac{\delta_{ij} (\not{p} + m_c)}{p^2 - m_c^2} - \frac{g_s G_{\mu\nu}^{ij}}{4} \frac{\sigma^{\mu\nu} (\not{p} + m_c) + (\not{p} + m_c) \sigma^{\mu\nu}}{(p^2 - m_c^2)^2} \nonumber \\
&+\frac{g^{2}}{12}G_{\alpha \beta }^{A}G_{A}^{\alpha \beta
}\delta_{ij}m_{c}\frac{k^{2}+m_{c}{\!\not\!{k}}}{(k^{2}-m_{c}^{2})^{4}}+\cdots\Bigg],
\end{align}
where \( G_{\mu\nu}^{ij} \) denotes the gluon field strength tensor and \( g_s \) is the strong coupling constant. The superscript "QCD" in Eqn. \ref{eq::PQCD1} indicates that the correlation function is calculated in terms of QCD degrees of freedom. Following the spectral integral definition of the correlation function as in the physical side, $\Pi^{\text{QCD}}(p, T)$ can also be expressed as a spectral integral as
\begin{equation}
	\Pi^{\text{QCD}}(p^2,T) = \int_{16m_c^2}^{\infty} \frac{\rho^{\text{QCD}}(s,T)}{s - p^2} \, ds + \text{subtractions},
\end{equation}
where
\begin{equation}
	\rho^{\text{QCD}}(s,T) = \frac{1}{\pi} \text{Im} \, \Pi^{\text{QCD}}(s),
\end{equation}
is the spectral density in terms of partonic degrees of freedoms. The correlation function can further be decomposed into perturbative and non perturbative contributions as
\begin{equation}
	\Pi^{\text{QCD}}(p, T) = \Pi^{\text{pert}}(p, T) + \Pi^{\text{nonpert}}(p, T),
\end{equation}
where these contributions are calculated with the help of operation product expansion (OPE). Analogously the spectral density can be written as
\begin{equation}
	\rho(s) = \rho^{\text{pert}}(s) + \rho^{\text{nonpert}}(s).
\end{equation}
where $\rho^{\text{nonpert}}$ contains contributions up to dimension four. After applying the Borel transformation, and the parton-hadron duality approximation, which assumes that the integrals of both the QCD and phenomenological sides from the temperature-dependent continuum threshold ($s(T)$) to infinity are equal and thus cancel. Hence, the QCD side of the correlation function can be expressed as 
\begin{align}
	\Pi^{\text{QCD}}(M^2, s(T), T) &= \int_{16m_c^2}^{s(T)} ds\, \rho^{\text{QCD}}(s, T) e^{-s/M^2} + \Pi^{\text{Dim4}}(M^2, s(T), T),
\end{align}
where $\Pi^{\text{Dim4}}(M^2, s(T), T)$ contains the nonperturbative contributions from dimension four condansates that can not be expressed in $\rho^{\text{nonpert}}$. The analytical expressions of the spectral densities and $\Pi^\text{Dim4}$, including the kinematic constraints, are provided in Appendix A, where $ \Pi^\text{Dim4}$ is computed separately. 

After following these traditional steps in QCDSR, the sum rule for the decay constant is written as
\begin{equation}\label{eq:decconSR}
	f^2 (T) = \frac{e^{m^2 (T)/M^2}}{m^2 (T)}\Pi(M^2, s(T), T),
\end{equation}
where $M^2$ is the Borel parameter, and $\Pi(M^2, s(T), T)$ is the Borel-transformed correlation function integrated up to the temperature dependent continuum threshold parameter. In Eqn. \ref{eq:decconSR}, $f(T)$ represents the decay constant and $m(T)$ the mass of the state at temperature $T$. Both quantities also implicitly depend on $s(T)$ and $M^2$, although these dependences are not shown expicitly for brevity.  

Using Eq.~(\ref{eq:decconSR}), the temperature-dependent mass sum rule is obtained as
\begin{equation}
	m(T) = \sqrt{\frac{\Pi' (M^2, s_0, T)}{\Pi(M^2, s_0, T)}},
\end{equation}
where $\Pi'(M^2, s_0, T) = \frac{d}{d(-1/M^2)} \Pi(M^2, s_0, T)$ represents the derivative of the Borel-transformed correlation function with respect to $-1/M^2$. This ensures that the extracted mass remains consistent with the sum rule formalism while minimizing unwanted contributions from higher excited states and the continuum. These expressions allow us to analyze the thermal behavior of the $X(6200)$ state and its structural stability at finite temperatures.

In TQCDSR, the nonperturbative effects are encoded in the thermal averages of gluon condensates. At finite temperature, the breaking of Lorentz invariance due to the thermal medium introduces a preferred frame defined by the four-velocity \( u^\mu \) of the hot medium. The thermal gluon condensate is then parameterized as \cite{Mallik:1997pq}
\begin{align}
\langle \text{Tr}^c G_{\alpha\beta} G_{\lambda\sigma} \rangle_T &= (g_{\alpha\lambda} g_{\beta\sigma} - g_{\alpha\sigma} g_{\beta\lambda}) C_1(T) \nonumber \\	&- (u_\alpha u_\lambda g_{\beta\sigma} - u_\alpha u_\sigma g_{\beta\lambda} - u_\beta u_\lambda g_{\alpha\sigma} + u_\beta u_\sigma g_{\alpha\lambda}) C_2(T),
\end{align}
where \( C_1(T) \) and \( C_2(T) \) are scalar functions that depend on the temperature and characterize the medium induced modifications of the gluon condensates and the gluonic part of the energy density, which is given by
\begin{align}
C_1(T) = \frac{1}{24} \left\langle G^a_{\mu\nu} G^{a\,\mu\nu} \right\rangle_T, \quad
C_2(T) = \frac{1}{6} \left\langle u^\lambda \Theta^g_{\lambda\sigma} u^\sigma \right\rangle_T.
\end{align}
In our calculations, we employ the temperature-dependent strong coupling $\alpha_s$ \cite{Kaczmarek:2004gv,Morita:2007hv}, with $\Lambda_{\overline{MS}}\simeq T_{c}/1.14$, where $T_c=155$ MeV~\cite{Andronic:2017pug}. The perturbative coupling of QCD is given by
\begin{eqnarray}\label{geks2T}
g_{pert}^{2}(T)=\frac{1}{\frac{11}{8\pi^2}\ln\Big(\frac{2\pi T}{\Lambda_{\overline{MS}}}\Big)+\frac{51}{88\pi^2}\ln\Big[2\ln\Big(\frac{2\pi
T}{\Lambda_{\overline{MS}}}\Big)\Big]},
\end{eqnarray}
where the coupling constant $g(T)$ is related to the perturbative one through relation $g^2(T)=2.096g_{\text{pert}}^2(T)$. This relation holds for temperatures above 100 MeV; for lower temperatures, we fix its value to those at $T=100$ MeV.
The temperature dependence of the continuum threshold $s(T)$ is parametrized as
\begin{eqnarray}
s(T) =s_0 \left[ 1 - \left( \frac{T}{T_c} \right)^8 \right] + 16m_c^2 \left( \frac{T}{T_c} \right)^8,
\end{eqnarray}
where $s_0$ represents the zero-temperature continuum threshold. This parametrization captures the essential physics of the system across different temperature regimes through distinct physical implications: In the low-temperature limit ($T \ll T_c$), the term $\left(\frac{T}{T_c}\right)^8$ becomes negligible, and $s(T)$ reduces to $ s_0$, maintaining the system's properties at zero temperature.
As the temperature approaches to $T_c$, a competition emerges between the suppression of $s_0$ term and the growing dominance of $16 m_c^2$ contribution, reflecting the onset of phase transition dynamics. At $T = T_c$, the first term vanishes completely, and the system is characterized solely by the term $16 m_c^2$, which arises from the mass of the c-quarks that constitute $X(6200)$. The temperature-dependent continuum threshold shows a direct correlation to the thermal light quark condensate in systems with light-light or heavy-light quarks.  In contrast, for heavy-heavy quark systems,this behavior deviates significantly, necessitating a separate treatment, as discussed in Refs~\cite{Dominguez:2009mk,Dominguez:2010mx}.

Following aforementioned discussions on the nature of $s(T)$, it is evident that obtained thermal sum rules for the mass and the decay constant should reproduce the same results the traditional sum rules at $T=0$. Therefore, they should satisfy standard reliability criteria, such as the pole contribution (PC), which is defined as the ratio of the ground state contribution to the full spectrum. A higher PC value generally indicates a more stable sum rule analysis, as it implies a dominant ground-state contribution. However, as the temperature increases, the PC typically decreases due to thermal effects, leading to a stronger continuum influence. Thus, when performing sum rule analyses, an appropriate PC range is selected to guarentee the stability and the accuracy of the results. For the \(\eta_c \eta_c\) molecular state, the analysis was conducted within a PC range of $ 86\% $ to $ 54\% $.

Additionally, in the TQCDSR framework, physical quantities such as the mass and decay constant should ideally be independent of the auxiliary parameters $M^2$ and $s_0$. These parameters are introduced to suppress contributions from the higher states and the continuum, ensuring that the extracted hadronic properties remain stable and physically meaningful. However, in practical applications, a residual dependence on these parameters is inevitable, necessitating a careful selection of their optimal range.  

\begin{figure}[htbp]
	\centering
	\begin{varwidth}{\linewidth}
		\includegraphics[totalheight=6cm,width=8cm]{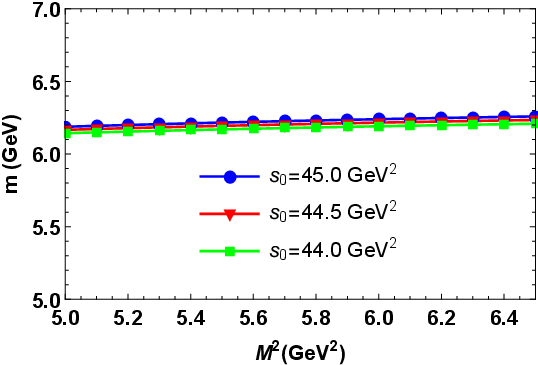}\,\, %
		\includegraphics[totalheight=6cm,width=8cm]{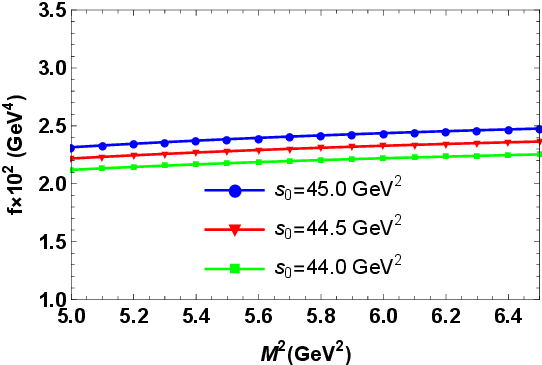}
		\caption{Stability analysis of the mass and decay constant of $X(6200)$ versus $M^2$ or different values of continuum thresholds $s_0$.}\label{fig:mfvsM2} 
	\end{varwidth}
\end{figure}

\begin{figure}[htbp]
	\centering
	\begin{varwidth}{\linewidth}
		\includegraphics[totalheight=6cm,width=8cm]{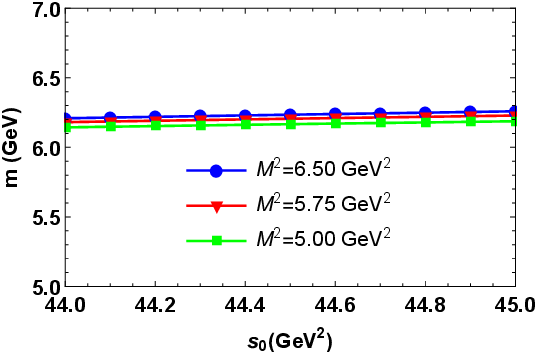}\,\, %
		\includegraphics[totalheight=6cm,width=8cm]{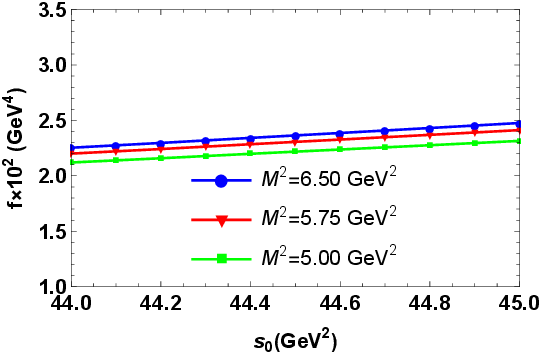}
		\caption{Dependence of the vacuum spectroscopic characteristics of the $X(6200)$ state on the continuum threshold parameter $s_0$, evaluated at selected $M^2$ values.} \label{fig:mfvss0}
	\end{varwidth}
\end{figure}

\begin{figure}[htbp]
	\centering
		\begin{varwidth}{\linewidth}
		\includegraphics[totalheight=6cm,width=8cm]{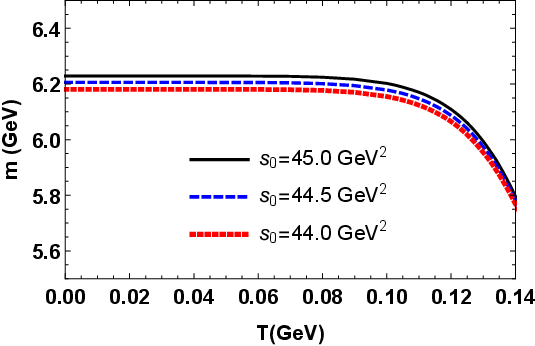}\,\, %
		\includegraphics[totalheight=6cm,width=8cm]{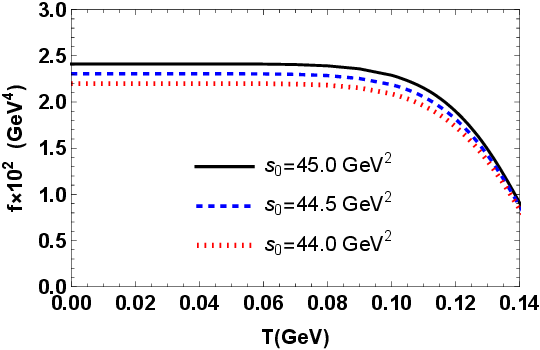}
		\caption{Dependence of the mass and decay constant of the \(X(6200)\) state on temperature, at fixed $M^2 = 5.75~\text{GeV}^2$ at selected values of $s_0$.
		} \label{fig:mfvsT}
\end{varwidth}
\end{figure}

To achieve reliable results, we follow the standard stability criteria, which involve minimizing the sensitivity of the extracted mass and decay constant to variations in $M^2$ and $s_0$ at $T=0$. By systematically analyzing the behavior of obtained sum rules under different values of these parameters, we determine their optimal working windows, as summarized in Table~\ref{tab:parameters}. To obtain the vacuum values of the exotic state $X(6200)$, we use the input data in given Table ~\ref{tab:numvalue}.
\begin{table}[h!]
	\centering
	\caption{The parameter ranges $ M^2 $ of $s_0$ and that ensure the stability of the sum rules.}
	\begin{tabular}{cc}
		\hline	\hline
		\textbf{Parameters} & \textbf{Range} \\
		\hline
		$M^2$ & $5 \, \text{GeV}^2 \leq M^2 \leq 6.5 \, \text{GeV}^2$ \\
		$s_0$ & $44 \, \text{GeV}^2 \leq s_0 \leq 45 \, \text{GeV}^2$ \\
		\hline	\hline
	\end{tabular}\label{tab:parameters}
\end{table}
\begin{table}[h!]
	\centering
	\caption{Values of the essential parameters employed in computational analysis.}
	\begin{tabular}{c c}
		\hline	\hline
		\textbf{Input parameters} & \textbf{Numeric Values} \\
		\hline
		$m_c$ & $1.27 \pm 0.02~\text{GeV}$ \cite{ParticleDataGroup:2024cfk} \\		
		$ m_{\eta_c} $&	$2984.1 \pm0.4~\text{MeV}$ \cite{ParticleDataGroup:2024cfk} \\
		$\langle 0 | \frac{\alpha_s G^2}{\pi} | 0 \rangle$ & $(0.0202 \pm 0.0011) \, \text{GeV}^4$ \cite{SVZ,Narison:2018dcr} \\
		\hline	\hline
		\end{tabular}\label{tab:numvalue}
\end{table}
The mass and decay constant dependencies on the Borel mass parameter for the $\eta_c\eta_c$ molecular configuration are illustrated in Figure~\ref{fig:mfvsM2}, plotted at zero temperature. Similarly, Figure~\ref{fig:mfvss0} depicts dependencies on the continuum threshold parameter, also evaluated at $T=0$. The thermal evolution of both mass and decay constant are presented in Figure~\ref{fig:mfvsT}, highlighting how they vary with increasing temperature.

The curves plotted in Figure~\ref{fig:mfvsT} for different fixed parameters show that the mass of the $ X(6200)$ state remains relatively stable at lower temperatures, then begins to decrease noticeably as the system approaches the critical temperature, particularly for $ T> 0.08~\mathrm{GeV} $. Similarly, the decay constant exhibits a significant decrease with increasing temperature, indicating that the decay constant of the $ {X}$ state is highly sensitive to thermal changes. This behavior suggests a reduction in the state’s binding strength as the temperature rises, consistent with the expected weakening of particle interactions and binding energy in a thermal medium.

These results underscore the importance of considering thermal effects into theoretical models of the $ X(6200)$ state. The observed sensitivity of its properties to temperature variations offers valuable insights for future experimental and theoretical investigations, enhancing our understanding of hadronic interactions in hot and dense environments.
\section{Discussion}\label{sec:conclusion}
In this work, we investigated the thermal behavior of the fully-charm exotic state, modeled as an $\eta_c \eta_c$ molecule with quantum numbers $J^{PC} = 0^{++}$, and determined the temperature dependence of its mass and decay constant using TQCDSR. Our results indicate a significant temperature dependence of these properties. As a benchmark, we summarize the mass and decay constant values at zero temperature and at $T = 0.14$ GeV, which is close to the critical temperature $T_c$, in Table~\ref{tab:mass_decay}.
\begin{table}[h!]
	\centering
	\begin{tabular}{ccc}
		\hline
		\textbf{T (GeV)} & \textbf{Mass (MeV)} & $\mathbf{f \times 10^{2}\ (\text{GeV}^4)}$ \\
		\hline
		$ 0$ & $6201.74 \pm 57.82$ & $2.31 \pm 0.18$ \\
		$0.14$ & $5782.43 \pm 37.42$ & $0.88 \pm 0.05$ \\
		\hline
	\end{tabular}
	\caption{Mass and decay constant of the $\eta_c, \eta_c$ molecule at $T = 0$ and $T = 0.14$ GeV.}
	\label{tab:mass_decay}
\end{table}

As seen from Table~\ref{tab:mass_decay}, the estimated mass and decay constant at zero temperature align well with existing literature, supporting the reliability of the obtained sum rules. Furthermore, at $T = 0.14$ GeV, the mass decreases by approximately $7\%$, while  decay constant drops about $62\%$. The rate of decrease in both quantities rises rapidly in the close vicinity of $T_c$.

In conclusion, our study using TQCDSR reveals significant temperature dependence in both mass and decay constant of the ${X}$ resonance. These findings provide valuable insights into particle behavior under extreme conditions, such as those found in the early universe or astrophysical plasmas. The observed thermal effects may also have implications for physics beyond the SM, , including scenarios involving extra dimensions, supersymmetry, or dark matter. Further targeted investigations are necessary to understand the underlying quantum dynamics governing this fully-charm exotic system.
\appendix
\section{Analytical Expressions of Temperature Dependent Spectral Densities}
In this appendix, we present the complete analytical expressions for the perturbative spectral density $\rho^{\text{pert}}$ and the nonperturbative spectral density $\rho^{\text{nonpert}}$, which includes contributions from dimension-4 gluon condensates and the gluonic component of the energy density. Additionally, we provide the analytical form of $\Pi^{\text{Dim4}}(M^2,T)$, representing the dimension-four contributions to the sum rules that are not expressed in $\rho^{\text{nonpert}}$. These expressions retain all kinematic dependencies and are given without any truncation or approximation, ensuring clarity and completeness of our analysis.

The perturbative spectral density $\rho^{\text{pert}}$ is obtained as
\begin{equation}
	\rho^{\text{pert}} (s,T)= -\frac{1}{2048\pi^6}\int_0^1 dr \int_0^{1-r} dw \int_0^{1-r-w} dz \frac{\mathcal{N}(s, r, w, z)}{ \mathcal{D}^8(r, w, z)}\theta[L(s, r, w, z)]
	\label{eq:rhoPert},
\end{equation}
where the numerator $ \mathcal{N} $ is defined as
\begin{align}
\nonumber	\mathcal{N} (s, r, w, z)&= (-\mathcal{D} m_c^2 + r s w z (-1 + r + w + z))^2 \Big[ -231 r^3 s^2 w^3 z^3 ( r + w + z-1)^3  \\
\nonumber	& - 2 \mathcal{D} m_c^2 rs w z (-1 + r + w + z)  \Big( 5 r w (-1 + r + w) \big( r^2 + (-1 + w) w + r \\
\nonumber	& \times (-1 + 13 w) \big) + (-1 + r+ w) \big( 5 r^3 + 5 (-1 + w) w^2 + 5 r^2 ( 3 w -1) + 3 r w   \\
\nonumber	&\times (-18 + 5 w) \big) z- \big( 50 r^3 + 5 (-1 + w) w (10 w-11) + 15 r^2 ( 10 w-7) + r   \\
\nonumber	&\times(55 + 2 w (-83 + 75 w)) \big) z^2- 110 (-1 + r + w) (r + w) z^3 - 55 (r + w) z^4 \Big)\\
\nonumber	&+ \mathcal{D}^2 m_c^4 \Big( 2 r w ( r + w-1) \big( 2 (-1 + w) w + r^2 (2 + 33 w) + r (-2 + w (-7  \big)\\
\nonumber	&  + 33 w)) + (-1 + r + w) \big( 4 (-1 + w) w^2 + 4 r^3 (1 + 33 w)+ 3 r w (1 - 40 w  \\
\nonumber	&+ 44 w^2) + 4 r^2 (-1 + 6 w (-5 + 11 w)) \big) z+ \big( 66 r^4 + 4 r^3 (-43 + 99 w)+ 2 (-1   \\
\nonumber	& + w) w (22 + w( 33 w-53))+ 6 r^2 (25 + 2 w (-54 + 55 w)) + r (-44 + w (311  \\
\nonumber	&  + 36 w(-18 + 11  w))) \big) z^2 + 44 (-1 + r + w) (r + w) (-2 + 3 r + 3 w) z^3 + 22   \\
	& \times(r+w) (-2 + 3 r  +  3 w) z^4 \Big) \Big],
\end{align}
and the common denominator $ (\mathcal{D}) $ is defined as
\begin{eqnarray}
	\mathcal{D}(r, w, z) &=& w z (w + z - 1) + r^2(w + z) + r(w + z)(w + z - 1).
\end{eqnarray}
In Eq.~(\ref{eq:rhoPert}), the function $\theta[L(s, r, w, z)]$ is the Heaviside step function, which ensures that the contributions to spectral density are considered within the kinematically allowed region defined by the condition $\theta[L(s, r, w, z)] > 0$. The quantity $ L(s, r, w, z) $ is described by
\begin{align}
	\nonumber	L(s, r, w, z) &= 
		\frac{
			\left(r^2 + r(z - 1 + w) + z(z - 1 + w)\right)
		}{
			\left(w z (z - 1 + w) + r^2 (w + z) 
			+ r \left(w^2 + (z - 1) z + w (2z - 1)\right) \right)^2
		} \\
	\nonumber	& \times \Big[ r s w z (-1 + r + w + z) 
		- m_c^2 \big( w z (z - 1 + w) 
		+ r^2 (w + z) \\
		& + r \left(w^2 + (z - 1) z + w (2z - 1) \right) \big) \Big],
\end{align}
which covers the kinematic constraints of the process, and its positivity ensures that the spectral density is evaluated only within the physical phase space.

The variables $r$, $w$ and $z$ are dimensionless Feynman parameters describing momentum fractions in the $\eta_c\eta_c$ molecular system. The charm quark mass $m_c$ appears as an explicit energy scale in the spectral function calculation. 

The nonperturbative part of the spectral density is calculated as
\begin{align}
\nonumber	\mathcal{\rho}^{\text{nonpert}}(s,T)
	&= \int_0^1 dr \int_0^{1 - r} dw \int_0^{1 - r - w} dz 
	\Bigg[ 
	\frac{1}{3072\, \mathcal{D}^5 \pi^4} \,\mathcal{H}(s, r, w, z) \, \left\langle \frac{\alpha_s G^2}{\pi} \right\rangle_T\nonumber\\
	&+ \frac{1}{256\, \mathcal{D}^6 \pi^6} \, \mathcal{Y}(s, r, w, z) \, g_s^2 (T) \, \left\langle u^\lambda \Theta^g_{\lambda\sigma} u^\sigma \right\rangle_T 
	\Bigg] \theta[L(s, r, w, z)],
\end{align}
where  $ 	\mathcal{H}(s, r, w, z) =\sum_{n=0}^{8} \mathcal{P}_n z^n $,
and $\mathcal{P}_n$ represents the coefficients of the powers of $z$. The $z$-power separated polynomials of $\mathcal{H}$ are described as
%
%
\begin{align}
\nonumber\mathcal{P}_0 (r,w) &= 2 \mathcal{D}^2 m_c^4 r w (1 - r - w) \Big[ 2 r^4 (11 - 9 w) + (1 - w) w \big\{ 1 + 2 (1 - w) w \big\} \\
\nonumber& + 2 r^3 \big\{ 22 + w (-31 + 87 w) \big\} + 3 r^2 \big\{ -7 + w (25 + 22 w + 58 w^2) \big\} \\
& + r \Big\{ -1 + w \big( -18 + w \big[ 27 + 2 w (-7 + 9 w) \big] \big) \Big\} \Big],
\end{align}
%
%
\begin{align}
\nonumber\mathcal{P}_1 (r,w)	&= \mathcal{D} m_c^2 (1 - r - w) \Big[
3 r^2 s w^2 (1 - r - w) \Big\{
(-1 + r) r (-1 + 22 (-1 + r) r) + w \\
\nonumber&+ r \big(18 + r ( 32 r- 63)\big) w - 3 \big(1 + r (5 + 82 r)\big) w^2 + 4 (1 - 4 r) w^3 - 2 w^4
\Big\}+ \mathcal{D} m_c^2  \\
\nonumber&\times \Big\{
-2 (-1 + r) r^2 (-1 + 22 (-1 + r) r) + r (-15 + 2 r (-30 + r (179 + 2 r (-91   \\
\nonumber&+ 18  r)))) w+ 2 \big(-1 + r [-6 + r ( 4 r-3) (-35 + 66 r)]\big) w^2 + 2 \{3 + r (35 - 242 r    \\
&+ 456 r^2)\}w^3+ 4 \big(-2 + r (-31 + 132 r)\big) w^4 + 4 (1 + 18 r) w^5\Big\}\Big], 
\end{align}
%
%
\begin{align}
\nonumber		\mathcal{P}_2 (r,w) &= \mathcal{D}\, m_c^2 \Big[3 r s w (-1 + r + w) \Big\{(-1 + r)^2 r^2 \big( -1 + 22 (-1 + r) r \big)+ (-1 + r) r \Big( 13  \\
\nonumber		& + r \big( -25 + 2 r \big( -67 + 77 r \big) \big) \Big) w + \Big( -1 + r \big( 62 + 3 r \big( -108 + r \big(  88 r - 1 \big) \big) \big) \Big)  \\
\nonumber		&\times w^2 + \Big( 4 + r \big( -107 + (285 - 98 r) r \big) \Big) w^3 + (-7 + 72 r) w^4 + 2 (3 - 7 r) w^5 \\
\nonumber		& - 2 w^6\Big\}+ \mathcal{D}\, m_c^2 \Big\{36 r^6 + 12 r^5 \big( -23 + 4 w \big) + 2 r^4 \Big( 271 + 2 w \big( -92 + 27 w \big) \Big) \\
\nonumber		&+ r \Big( -1 + 2 w \Big) \Big( 22 + w \big( -71 + 4 w \big( -1 + 2 w \big( 7 + 3 w \big) \big) \big) \Big) + 4 r^3 \Big( -105   \\
\nonumber		&+ w \big( 119 + 2 w \big( 59 + 24 w \big) \big) \Big)+ 2 r^2 \Big( 70 + w \big( -129 + w \big(   380 w + 54 w^2-279 \big) \big) \Big) \\
		&+ 2 (-1 + w) w \Big( 11 + w \big( -59 + w \big( 79 + 6 w \big( -8 + 3 w \big) \big) \big) \Big)\Big\}	\Big],
	\end{align}
%
%
	\begin{align}
\nonumber	\mathcal{P}_3 (r,w) &=- (-1 + r + w) \Bigg[
			120 r^3 s^2 w^3 (-1 + r + w)^2 \big( r^2 + (-1 + w) w + r (-1 + 13 w) \big) \\
\nonumber			&+ 4 \mathcal{D}^2 m_c^4 \Big\{
			42 r^4 + r^3 (-173 + 288 w) + r^2 \big( 253 + w (-809 + 492 w) \big) \\
\nonumber			&+ w \big( -110 + w (253 + w (-197 + 42 w)) \big) + r \big( -110 + w (583 + w (-833  \\
\nonumber			&+ 288 w)) \big)
			\Big\}- 3 \mathcal{D} m_c^2 r s w \Big\{
			100 r^5 + 3 r^4 (-81 + 134 w) + 15 r^3 (13 - 58 w + 34 w^2) \\
\nonumber			&+ r^2 \big( -63 + w (823 - 478 w + 270 w^2) \big) + (-1 + w) w \big( -11 + w (52 + w (-47  \\
			&+ 4 w)) \big)+ r \big( 11 + w (-346 + w (727 + 6 w (-73 + 11 w))) \big)
			\Big\}
			\Bigg],
		\end{align}
%
%
\begin{align}
\nonumber	\mathcal{P}_4(r, w) &= 120\,r^3 s^2 w^3 (r + w - 1)^2 (8r^2 - 19r + 11 - 17r w - 19w + 8w^2) \\
\nonumber	& - 2\,\mathcal{D}^2 m_c^4 \Big[342r^4 + 2r^3 (744w - 695) + (w - 1)w (171w^2 - 524w + 715) \\
\nonumber	& + r^2 (2292w^2 - 4436w + 1763) + r (1488w^2 - 4436w + 3603)w - 715\Big] \\
\nonumber	& - 3\,\mathcal{D} m_c^2 r s w \Big[76r^5 + r^4 (578w - 507) + 4r^3 (75w^2 - 88w + 249)+ (w - 1)w \\
\nonumber	& \times (196w^3 - 551w^2 + 565w - 209) + 2r^2 (870w^3 - 1507w^2 + 898w - 387) \\
	& + r (938w^3 - 2240w^2 + 1916w - 833) + 209\Big],
\end{align}
%
%
\begin{align}
\nonumber	\mathcal{P}_5(r,w) &= -3 (r + w - 1) \Big[4 \mathcal{D}^2 m_c^4 (r + w) (60r + 60w - 121) - 40 r^3 s^2 w^3 (41r^2 - 85r \\
\nonumber &+ 44 + 49r w - 85w + 41w^2) +\mathcal{D} m_c^2 r s w \big(848r^3 + r^2 (2670w - 1807)\\
	& + r (2718w^2 - 2888w + 935) + w (896w^2 - 1807w + 935)\big)\Big],
\end{align}
%
%
\begin{align}
\nonumber	\mathcal{P}_6 (r,w)&= -4 \mathcal{D}^2 m_c^4 (r + w) \Big[60r + 60w - 121\Big] + 120 r^3 s^2 w^3 \Big[65r^2 + (w - 1)(65w - 66) \\
\nonumber	&+ r (119w - 131)\Big] - 3 \mathcal{D} m_c^2 r s w \Big[1420r^3 + (w - 1)w (1420w - 1441)  \\
	&+ r^2 (4286w - 2861)+ r (4286w^2 - 5480w + 1441)\Big],
\end{align}
%
%
\begin{align}
\mathcal{P}_7 (r,w)&= -264 r s w (r + w - 1) \big[-20 r^2 s w^2 + 11 \mathcal{D} m_c^2 (r + w)\big], \\
\mathcal{P}_8 (r,w)&= -66 r s w \big[-20 r^2 s w^2 + 11 \mathcal{D} m_c^2 (r + w)\big],
\end{align}
and
\begin{align}
\nonumber	\mathcal{Y}(s, r, w, z)&= r\, s\, w\, z\, (1 - r - w - z) \Big[5\, r^2\, s\, w^2\, z^2\, (1 - r - w - z)^2\,
	(-r + r^2 - w + 13 r w  \\ 
\nonumber	 &+ w^2- 11 (1 - r - w) z- 11 z^2) 
	+2\mathcal{D} m_c^2 \Big( + (1 - w) w^2 (1 - r - w)^2 (r + w) \\
\nonumber	&+ w^2 (1 - r - w)\, (r + 7 r^2 (1 + 2 r) + w + 6 r (-1 + 3 r) w + (-2 + 5 r) w^2 + w^3) z \\
\nonumber	&+ \big[(1 - w)^2 w^2 (1 - 3 w) + r^4 (-1 + 14 w) + r (1 - w) w (24 + w (-42 + 23 w)) \\
\nonumber	&+ r^3 (2 + w (-53 + 56 w)) + r^2 (-1 + w (63 + w (-99 + 62 w)))\big] z^2 \\
\nonumber	&+ (1 - r - w) (r^3 + w^2 (-2 + 3 w) + r^2 (-2 + 27 w) + r w (-48 + 29 w)) z^3 \\
	&+ (r^3 + (1 - w) w^2 + 2 r w (-12 + 7 w) + r^2 (-1 + 14 w)) z^4 \Big) \Big].
\end{align}

The expression for $\Pi^{\text{Dim4}}(M^2, s(T),T)$, which governs nonperturbative dimension-four contributions that are not expressible in the nonperturbative spectral density, is calculated as
\begin{align}
\nonumber \Pi^{\text{Dim4}}(M^2, s(T), T) &= -\int_{16m_c^2}^{s(T)} ds \int_0^1 dr \int_0^{1 - r} dw \int_0^{1 - r - w} dz \\
\nonumber &\times \frac{
    \exp\left[-\frac{m_c^2}{M^2}\left(\frac{1}{r} + \frac{1}{w} + \frac{1}{z} - \frac{1}{\mathcal{R}}\right)\right] 
    \cdot m_c^6 \cdot \mathcal{P}(r,w,z)
}{ 
    6144\pi^4 r w z \cdot \mathcal{R}(r,w,z) \cdot \mathcal{Q}^3(r,w,z)
} \\
& \times \left\langle \frac{\alpha_s G^2}{\pi} \right\rangle_T \theta[L(s, r, w, z)]
\end{align}
where
\begin{equation}
\mathcal{R} =r + w + z - 1,	
\end{equation}
\begin{align}
	\mathcal{P} (r,w,z)&= 	r^2 w^2 (r + w - 1)^2 
	\left(10 r^2 + 22 r (w - 1) + 2 w (5 w - 11) + 11\right) \nonumber\\
	& + r w (r + w - 1) \Big[	20 r^4 	+ r^3 (82 w - 40)	+ 4 r^2 (w - 1)(31 w + 1)\nonumber\\
	& + 2 r (41 w^3 - 60 w^2 + 6 w + 12)+ 4 (w - 1) w (5 w^2 - 10 w + 6)- 11 \Big]  \nonumber\\
	&\times  z+\Big[10 r^6 + 6 r^5 (13 w - 3)	+ r^4 (2 w - 1)(112 w - 5)+4 r^3 (78 w^3  \nonumber\\
    & - 69 w^2- 4 w + 1)+ r^2 (2 w + 1)(56 w^3 - 97 w^2 + 86 w - 1)+ r w \nonumber\\
	& \times (39 w^4 - 61 w^3 - 8 w^2 + 84 w - 35)	+ (w - 1)^2 w^2 (10 w^2 + 2 w- 1) \Big]   \nonumber\\
	& \times z^2+2 (r + w - 1) \Big[	9 r^4+ r^3 (34 w + 4)+ 2 r^2 (25 w^2 + 8 w - 1) \nonumber\\
	& + 2 r w (17 w^2 + 8 w - 13)+ w^2 (9 w^2 + 4 w - 2) \Big] z^3+ \Big[4 r^4	+ 14 r^3 (w + 1)\nonumber\\
	& +  r^2 (20 w^2 + 46 w - 7) + 2 r w (7 w^2 + 23 w - 18)+ w^2 (4 w^2 + 14 w - 7) \Big]  \nonumber\\
	& \times z^4 - 6 (r + w - 1) (r + w)^2 z^5- 2 (r + w)^2 z^6,
\end{align}
and
\begin{align}
\mathcal{Q}(r,w,z) &= w z (w + z - 1) + r^2 (w + z) + r (w + z - 1)(w + z).
\end{align}
\section*{References}


\begin{thebibliography}{99}
	%
	\bibitem{Godfrey:2008nc}
	S.~Godfrey and S.~L.~Olsen,
	\href{https://www.annualreviews.org/content/journals/10.1146/annurev.nucl.58.110707.171145}{Ann. Rev. Nucl. Part. Sci. \textbf{58}, 51-73 (2008)}
	%
	\bibitem{Guo:2017jvc}
	F.~K.~Guo, C.~Hanhart, U.~G.~Mei\ss{}ner, Q.~Wang, Q.~Zhao and B.~S.~Zou,
    \href{https://doi.org/10.1103/RevModPhys.90.015004}{Rev. Mod. Phys. \textbf{90}, no.1, 015004 (2018)
	[erratum: Rev. Mod. Phys. \textbf{94}, no.2, 029901 (2022)]}
	%
	\bibitem{Chen:2022asf}
	H.~X.~Chen, W.~Chen, X.~Liu, Y.~R.~Liu and S.~L.~Zhu,
	\href{https://iopscience.iop.org/article/10.1088/1361-6633/aca3b6}{Rept. Prog. Phys. \textbf{86}, no.2, 026201 (2023)}
	%
	\bibitem{Brambilla:2019esw}
	N.~Brambilla, S.~Eidelman, C.~Hanhart, A.~Nefediev, C.~P.~Shen, C.~E.~Thomas, A.~Vairo and C.~Z.~Yuan,
	\href{https://www.sciencedirect.com/science/article/abs/pii/S0370157320301915?viab}{Phys. Rept. \textbf{873}, 1-154 (2020)}
	%
	\bibitem{Albuquerque:2018jkn}
	R.~M.~Albuquerque, J.~M.~Dias, K.~P.~Khemchandani, A.~Mart\'\i{}nez Torres, F.~S.~Navarra, M.~Nielsen and C.~M.~Zanetti,
	\href{https://iopscience.iop.org/article/10.1088/1361-6471/ab2678}{J. Phys. G \textbf{46}, no.9, 093002 (2019)}
	%
	\bibitem{Ali:2017jda}
	A.~Ali, J.~S.~Lange and S.~Stone,
	\href{https://linkinghub.elsevier.com/retrieve/pii/S0146641017300716}{Prog. Part. Nucl. Phys. \textbf{97}, 123-198 (2017)}
	%
	\bibitem{LHCb:2020bwg}
	R. Aaij et al. (LHCb Collaboration), 
	\href{https://www.sciencedirect.com/science/article/pii/S2095927320305685?via}{Sci. Bull. \textbf{65}, 1983 (2020)}
\bibitem{Gell-Mann:1964ewy}
M.~Gell-Mann,
\href{https://doi.org/10.1016/S0031-9163(64)92001-3}{Phys. Lett. \textbf{8}, 214 (1964)}
	\bibitem{Iwasaki:1975pv}
	Y.~Iwasaki,
	\href{https://academic.oup.com/ptp/article/54/2/492/1831244?login=false}{Prog. Theor. Phys. \textbf{54}, 492 (1975)}
	%
	\bibitem{Nefediev:2021pww}
	A.~V.~Nefediev,
	\href{https://link.springer.com/article/10.1140/epjc/s10052-021-09511-z}{Eur. Phys. J. C \textbf{81}, no.8, 692 (2021)}
	%
	\bibitem{Faustov:2022mvs}
	R.~N.~Faustov, V.~O.~Galkin and E.~M.~Savchenko,
	\href{https://www.mdpi.com/2073-8994/14/12/2504}{Symmetry \textbf{14}, no.12, 2504 (2022)}
	%
	\bibitem{Dong:2020nwy}
	X.~K.~Dong, V.~Baru, F.~K.~Guo, C.~Hanhart and A.~Nefediev,
	\href{https://journals.aps.org/prl/abstract/10.1103/PhysRevLett.126.132001}{Phys. Rev. Lett. \textbf{126}, no.13, 132001 (2021)}
	%
	\bibitem{Song:2024ykq}
	Y.~L.~Song, Y.~Zhang, V.~Baru, F.~K.~Guo, C.~Hanhart and A.~Nefediev,
	\href{https://journals.aps.org/prd/abstract/10.1103/PhysRevD.111.034038}{Phys. Rev. D \textbf{111}, no.3, 034038 (2025)}
	%
	\bibitem{Agaev:2023ruu}
	S.~S.~Agaev, K.~Azizi, B.~Barsbay and H.~Sundu,
	\href{https://link.springer.com/article/10.1140/epjp/s13360-023-04562-5}{Eur. Phys. J. Plus \textbf{138}, no.10, 935 (2023)}
	%
	\bibitem{Shuryak:1980tp}
	E.~V.~Shuryak,
	\href{https://linkinghub.elsevier.com/retrieve/pii/0370157380901052}{Phys. Rept. \textbf{61}, 71-158 (1980)}
	%
	\bibitem{Harris:2023tti}
	J.~W.~Harris and B.~M\"uller,
	\href{https://link.springer.com/article/10.1140/epjc/s10052-024-12533-y}{Eur. Phys. J. C \textbf{84}, no.3, 247 (2024)}
	%
	\bibitem{Krintiras:2024qzx}
	G.~K.~Krintiras [ATLAS and CMS],
	\href{https://www.sif.it/riviste/sif/ncc/econtents/2024/047/04/article/5}{Nuovo Cim. C \textbf{47}, no.4, 148 (2024)}
	%
	\bibitem{Stoecker:2004qu}
	H.~Stoecker,
	\href{https://linkinghub.elsevier.com/retrieve/pii/S0375947404013028}{Nucl. Phys. A \textbf{750}, 121-147 (2005)}
	%
	\bibitem{Elfner:2022iae}
	H.~Elfner and B.~M\"uller,
	\href{https://iopscience.iop.org/article/10.1088/1361-6471/ace824}{J. Phys. G \textbf{50}, no.10, 103001 (2023)}
	%
	\bibitem{Prozorova:2024oal}
	V.~Prozorova [STAR],
	\href{https://pos.sissa.it/469/175}{PoS \textbf{DIS2024}, 175 (2025)}
	%
	\bibitem{Torres:2024ile}
	V.~V.~Torres [ALICE],
	\href{https://pos.sissa.it/476/612}{PoS \textbf{ICHEP2024}, 612 (2025)}
	%
	\bibitem{Fukushima:2025ujk}
	K.~Fukushima,
	\href{https://arxiv.org/pdf/2501.01907}{[arXiv:2501.01907 [hep-ph]]}.
	%
	\bibitem{Zhao:2023ucp}
	J.~Zhao, J.~Aichelin, P.~B.~Gossiaux and K.~Werner,
	\href{https://journals.aps.org/prd/abstract/10.1103/PhysRevD.109.054011}{Phys. Rev. D \textbf{109}, no.5, 054011 (2024)}
	
	\bibitem{Mishra:2023uhx}
	A.~Mishra, A.~Kumar and S.~P.~Misra,
	\href{https://journals.aps.org/prd/abstract/10.1103/PhysRevD.110.014003}{Phys. Rev. D \textbf{110}, no.1, 014003 (2024)}
	
	\bibitem{Sungu:2020zvk}
	J.~Y.~S\"ung\"u, A.~T\"urkan, H.~Sundu and E.~V.~Veliev,
	\href{https://link.springer.com/article/10.1140/epjc/s10052-022-10305-0}{Eur. Phys. J. C \textbf{82}, no.5, 453 (2022)}
	
	\bibitem{Sungu:2019ybf}
	J.~Y.~S\"ung\"u, A.~T\"urkan and E.~Veli Veliev,
	\href{https://www.actaphys.uj.edu.pl/index_n.php?I=R&V=50&N=9#1501}{Acta Phys. Polon. B \textbf{50}, 1501 (2019)}
	
	\bibitem{Turkan:2019anj}
	A.~T\"urkan, H.~Da\u{g}, J.~Y.~S\"ung\"u and E.~Veli Veliev,
	\href{https://iopscience.iop.org/article/10.1209/0295-5075/126/51001}{EPL \textbf{126}, no.5, 51001 (2019)}
	
	\bibitem{Gungor:2023ksu}
	E.~G\"ung\"or, H.~Sundu, J.~Y.~S\"ung\"u and E.~V.~Veliev,
	\href{https://link.springer.com/article/10.1007/s00601-023-01807-y}{Few Body Syst. \textbf{64}, no.3, 53 (2023)}
	
	\bibitem{Sungu:2020azn}
	J.~Y.~S\"ung\"u, A.~T\"urkan, E.~Sertbakan and E.~V.~Veliev,
	\href{https://link.springer.com/article/10.1140/epjc/s10052-020-08439-0}{Eur. Phys. J. C \textbf{80}, no.10, 943 (2020)}
	
	\bibitem{Azizi:2020itk}
	K.~Azizi and N.~Er,
	\href{https://doi.org/10.1103/PhysRevD.101.074037}{Phys. Rev. D \textbf{101}, no.7, 074037 (2020)}
	
	\bibitem{Azizi:2020yhs}
	K.~Azizi and N.~Er,
	\href{https://doi.org/10.1016/j.physletb.2020.135979}{Phys. Lett. B \textbf{811}, 135979 (2020)}
	
	\bibitem{Sungu:2024oax}
	J.~Y.~S\"ung\"u and N.~Er,
	\href{https://iopscience.iop.org/article/10.1088/1361-6471/ad66eb}{J. Phys. G \textbf{51}, no.12, 125001 (2024)}
	
	\bibitem{Veliev:2014tca}
	E.~V.~Veliev, K.~Azizi, H.~Sundu and G.~Kaya,
	\href{https://rjp.nipne.ro/2014_59_1-2/RomJPhys.59.p140.pdf}{Rom. J. Phys. \textbf{59-63}, no.1-2, 140 (2014)}
	
	\bibitem{Veliev:2011kq}
	E.~V.~Veliev, K.~Azizi, H.~Sundu, G.~Kaya and A.~T\"urkan,
	\href{https://link.springer.com/article/10.1140/epja/i2011-11110-8}{Eur. Phys. J. A \textbf{47}, 110 (2011)}
	
	\bibitem{Bozkir:2022lyk}
	G.~Bozk\i{}r, A.~T\"urkan and K.~Azizi,
	\href{https://link.springer.com/article/10.1140/epja/s10050-023-01187-1}{Eur. Phys. J. A \textbf{59}, no.11, 267 (2023)}
	
	\bibitem{Er:2022cxx}
	N.~Er and K.~Azizi,
	\href{https://link.springer.com/article/10.1140/epjc/s10052-022-10333-w}{Eur. Phys. J. C \textbf{82}, no.5, 397 (2022)}
	%
	\bibitem{SVZ} M. A. Shifman, A. I. Vainshtein, and V. I. Zakharov, 
	\href{https://doi.org/10.1016/0550-3213(79)90022-1}{Nucl. Phys. B \textbf{147}, 385 (1979)}
	%
	\bibitem{Bochkarev:1985ex} A. I. Bochkarev and M. E. Shaposhnikov,
	\href{https://doi.org/10.1016/0550-3213(86)90209-9}{Nucl. Phys. B \textbf{268}, 220 (1986)}
	%
	\bibitem{Andronic:2017pug}
	A.~Andronic, P.~Braun-Munzinger, K.~Redlich and J.~Stachel,
	\href{https://www.nature.com/articles/s41586-018-0491-6}{Nature {\bf 561}, no. 7723, 321 (2018)}
	%
	\bibitem{Aoki:2006br}
	Y.~Aoki, Z.~Fodor, S.~D.~Katz and K.~K.~Szabo,
	\href{https://linkinghub.elsevier.com/retrieve/pii/S0370269306012755}{Phys.\ Lett.\ B {\bf 643}, 46 (2006)}
	%
	\bibitem{Steinbrecher:2018phh}
	P.~Steinbrecher, [HotQCD Collaboration],
	\href{https://www.sciencedirect.com/science/article/pii/S0375947418301696?via%3Dihub}{Nucl.\ Phys. \ A {\bf 982}, 847 (2019)}
	%
	\bibitem{Fischer:2018sdj}
	C.~S.~Fischer,
	\href{https://linkinghub.elsevier.com/retrieve/pii/S014664101930002X}{Prog. Part. Nucl. Phys. \textbf{105},  1-60 (2019)}
	%
	\bibitem{Mallik:1997pq}
	S.~Mallik,
	Phys. Lett. B \textbf{416}, 373-378 (1998)
	%
	\bibitem{Kaczmarek:2004gv}
	O.~Kaczmarek, F.~Karsch, F.~Zantow and P.~Petreczky,
	\href{https://journals.aps.org/prd/abstract/10.1103/PhysRevD.70.074505}{Phys. Rev. D \textbf{70}, 074505 (2004) [erratum: Phys. Rev. D \textbf{72}, 059903 (2005)]}
	
	\bibitem{Morita:2007hv}
	K.~Morita and S.~H.~Lee,
	\href{https://journals.aps.org/prc/abstract/10.1103/PhysRevC.77.064904}{Phys. Rev. C \textbf{77}, 064904 (2008)}
	
	\bibitem{Dominguez:2009mk}
	C.~A.~Dominguez, M.~Loewe, J.~C.~Rojas and Y.~Zhang,
	Phys. Rev. D \textbf{81}, 014007 (2010)
	
	\bibitem{Dominguez:2010mx}
	C.~A.~Dominguez, M.~Loewe, J.~C.~Rojas and Y.~Zhang,
	Phys. Rev. D \textbf{83}, 034033 (2011)
	
	\bibitem{ParticleDataGroup:2024cfk}
	S.~Navas \textit{et al.} [Particle Data Group],
	\href{https://journals.aps.org/prd/abstract/10.1103/PhysRevD.110.030001}{Phys. Rev. D \textbf{110}, no.3, 030001 (2024)}
	
	\bibitem{Narison:2018dcr}
	S.~Narison,
	\href{https://www.worldscientific.com/doi/abs/10.1142/S0217751X18500458}{Int. J. Mod. Phys. A \textbf{33}, no.10, 1850045 (2018)}
	%
	\end{thebibliography}
\end{document}